\documentclass[journal]{IEEEtran}
\usepackage{amsmath,amsfonts}
\usepackage{amsthm}
\newtheorem{remark}{Remark}
\usepackage{array}

\usepackage[noadjust]{cite}
\usepackage{url}
\usepackage[font=footnotesize]{caption}
\usepackage{graphicx}
\usepackage{color}
\def\BibTeX{{\rm B\kern-.05em{\sc i\kern-.025em b}\kern-.08em
		T\kern-.1667em\lower.7ex\hbox{E}\kern-.125emX}}

\usepackage{tikz}
\newcommand\copyrighttext{%
	\footnotesize \copyright 2024 IEEE. Personal use of this material is permitted. Permission from IEEE must be obtained for all other uses, in any current or future media, including reprinting/republishing this material for advertising or promotional purposes, creating new collective works, for resale or redistribution to servers or lists, or reuse of any copyrighted component of this work in other works.}
\newcommand\copyrightnotice{%
	\begin{tikzpicture}[remember picture,overlay]
		\node[anchor=south,yshift=2pt] at (current page.south) {\fbox{\parbox{\dimexpr\textwidth-\fboxsep-\fboxrule\relax}{\copyrighttext}}};
	\end{tikzpicture}%
}

\newcommand{\secref}[1]{Sec. \ref{#1}}
\newcommand{\figref}[1]{Fig. \ref{#1}}
\newcommand{\tblref}[1]{Table \ref{#1}}
\newcommand{\cc}[1]{\mathcal{#1}}
\newcommand{\Nrep}{N_\mathrm{rep}}
\newcommand{\Tcycle}{T_\mathrm{cycle}}
\newcommand{\Dcycle}{D_\mathrm{cycle}}
\newcommand{\Rs}{R_\mathrm{s}}
\newcommand{\Tp}{T_\mathrm{p}}
\newcommand{\TR}{T_\mathrm{R}}
\newcommand{\Rdrop}{R_\mathrm{drop}}
\newcommand{\Dinv}{D_\mathrm{inv}}
\newcommand{\Dslot}{D_\mathrm{slot}}

\begin{document}

	\title{Analyzing the Scalability of Bi-static Backscatter Networks for Large Scale Applications}
	\author{Kartik Patel$^{*}$, Junbo Zhang$^{*}$, John Kimionis, Lefteris Kampianakis, Michael S. Eggleston, Jinfeng Du
		\thanks{
			This article has been accepted for publication in IEEE Journal of Radio Frequency Identification. This is the author's version which has not been fully edited and content may change prior to final publication. Citation information: DOI 10.1109/JRFID.2024.3514454.
			
			A part of this paper was presented as a poster at the 18th IEEE RFID conference between 4-6 June, 2024 in Boston, MA \cite{RFIDConf24MAC}. 
			
			$^*$ These authors contributed equally to this article.
			
			Junbo Zhang is with Carnegie Mellon University, Pittsburgh, PA 15213 USA (e-mail: junboz2@alumni.cmu.edu). 
			
			Kartik Patel, John Kimionis, Lefteris Kampianakis, Michael S. Eggleston, and Jinfeng Du are with Nokia Bell Labs, Murray Hill, NJ 07974 USA (e-mail: \{kartik.patel, ioannis.kimionis, lefteris.kampianakis, michael.eggleston, jinfeng.du\} @nokia-bell-labs.com).
	}}

	\maketitle
	\copyrightnotice

	\begin{abstract}
		Backscatter radio is a promising technology for low-cost and low-power Internet-of-Things (IoT) networks. The conventional monostatic backscatter radio is constrained by its limited communication range, which restricts its utility in wide-area applications. An alternative bi-static backscatter radio architecture, characterized by a dis-aggregated illuminator and receiver, can provide enhanced coverage and, thus, can support wide-area applications. 
		In this paper, we analyze the scalability of the bi-static backscatter radio for large-scale wide-area IoT networks consisting of a large number of unsynchronized, receiver-less tags. 
		We introduce the Tag Drop Rate (TDR) as a measure of reliability and develop a theoretical framework to estimate TDR in terms of the network parameters. We show that under certain approximations, a small-scale prototype can emulate a large-scale network. We then use the measurements from experimental prototypes of bi-static backscatter networks (BNs) to refine the theoretical model. Finally, based on the insights derived from the theoretical model and the experimental measurements, we describe a systematic methodology for tuning the network parameters and identifying the physical layer design requirements for the reliable operation of large-scale bi-static BNs. Our analysis shows that even with a modest physical layer requirement of bit error rate (BER) 0.2, 1000 receiver-less tags can be supported with 99.9\% reliability. This demonstrates the feasibility of bi-static BNs for large-scale wide-area IoT applications. 
	\end{abstract}
	

	\section{Introduction}
	\IEEEPARstart{T}{he} total cost of ownership of a large-scale dense IoT network is heavily dominated by the cost of the IoT devices. 
	Backscatter radio is a promising technology for such IoT networks as it enables low-cost end-nodes with long battery life~\cite{vannucci_software-defined_2008,van_huynh_ambient_2018,kampianakis_wireless_2014,jiangBackscatterCommunicationMeets2023}.
	Backscatter radio-based devices modulate the existing radio frequency (RF) carriers by modulating the load on their antenna. This results in an ultralow complexity passive device that offers orders of magnitude lower cost and power compared to conventional active radio technologies such as Wi-Fi, LTE, or Bluetooth. Therefore, backscatter radio is an ideal choice for massive-scale IoT network applications such as high-volume asset tracking, environmental and agricultural sensing applications, and security systems. 
	
	Radio frequency identification (RFID) systems are a common form of backscatter radio and have been used extensively in industrial and commercial settings for asset tracking of retail items, packaged goods, or pallets for several decades~\cite{chawla_overview_2007}. RFID, however, is limited by the short communication range between the tags and the reader. The reader is typically a monostatic device, i.e., the transmit and receive chains reside in the same box, which has two effects: (1) the tag's SNR is affected by the round-trip path loss (Tx$\rightarrow$Tag$\rightarrow$Rx) and (2) the overall operating range is governed by the forward link (or downlink), as opposed to the reverse link (or uplink). This is because a tag's power harvester and receiver have limited sensitivity since they are implemented as envelope detectors~\cite{daniel_dobkin_rf_2012}. Even though readers have much higher sensitivity levels afforded by their low noise amplifiers (LNA) and baseband gain stages, it does not contribute to a longer feasible range~\cite{wang_link_2008,griffin_complete_2009}. Therefore, more readers are required to increase the coverage area, which introduces a significant cost for network deployment in wide-area applications. 
	
	Prior work has shown that decoupling the receiver from the transmitter (\textit{bi-static architecture}) and creating distributed reader units can increase the operating range by one to two orders of magnitude~\cite{kimionis_increased_2014,vougioukas_could_2016}. Hence, the bi-static architecture significantly boosts the coverage of low-complexity backscatter radio systems and, as a result, supports wide-area applications. However, the network scalability of bi-static BNs, i.e., whether they can support large-scale network applications, is unclear. 
	
	Scalability for the monostatic systems is achieved through multiple access mechanisms that use synchronized readouts of thousands of sensors/tags. For example, the prevailing medium access control (MAC) protocol for RFID tags, EPC UHF Gen 2~\cite{global_epc_epc_2008}, uses a form of framed slotted ALOHA where the reader coordinates tags to multiplex them in time slots within a frame (inventory round). Such techniques, however, not only require that tags employ receivers that can process the reader's query commands, but also impose strict timing requirements between the transmitting and receiving parts of the reader. These restrictions reduce the area coverage of a bi-static system and introduce significant synchronization challenges since the illuminators and receiver units in a bi-static system are not inherently synchronized in time or frequency due to their remote deployments. 
	
	In this paper, we provide a framework for assessing the scalability of a bi-static BN consisting of an unsynchronized illuminator and receiver that read out a large number of receiver-less tags. 
	We first theoretically model the reliability of the bi-static BNs that use a random time division multiplexing as the MAC protocol. 
	We introduce a \textit{collision zone parameter} in the model which measures the effect of the collision of packets on their corruption. 
	We then show that with some approximations, the reliability of a large-scale network can be extrapolated from the measurements on a small-scale BN. 
	This allows us to estimate the collision zone parameter using a small-scale experimental testbed. Subsequently, we use the estimated collision zone parameter to design the parameters for a large-scale bi-static BN. 
	As a result, we show how a large-scale bi-static BN can achieve reliable performance (99.9\% reliability) for large-scale applications containing 1000 tags with modest physical layer requirements. These results, combined with the better coverage and lower deployment cost of a bi-static architecture, make it an attractive alternative to traditional monostatic architectures for wide-area large-scale passive IoT networks.
	
	\subsection{Related Work}
	
	Several implementations of BNs have demonstrated increased coverage using receiver-less tags with bi-static~\cite{kimionis_increased_2014, vougioukas_could_2016, alevizosMultistaticScatterRadio2018} or ambient architectures \cite{buttAmbientIoTMissing2024, alevizos_batteryless_2023, zhang_hitchhike_2016, zhang_freerider_2017, kellogg_passive_2016, kellogg_wi-fi_2014, talla_lora_2017, bharadia_backfi_2015,yuanEnablingNativeWiFi2023}. Some works have also introduced MIMO capabilities at both the illuminators and receivers to reduce direct link interference in bi-static architecture~\cite{kaplan_direct_2024, kaplan_dynamic_2022}. In this work, we evaluate the scalability of such bi-static BNs with uncoordinated illuminators and receivers.

	Although many previous studies have explored multiple access strategies for BNs~\cite{global_epc_epc_2008, hessar_netscatter_2019, mishraMultiTagBackscatteringMIMO2019, mishraSumThroughputMaximization2019a, alim_backscatter_2017, choi_devices_2021, cao_distributed_2020, ma_design_2019, kwon_dominant_2016, bharadia_backfi_2015, zhang_freerider_2017, alevizosMultistaticScatterRadio2018}, they often assume either monostatic architectures at the reader~\cite{global_epc_epc_2008, hessar_netscatter_2019, mishraMultiTagBackscatteringMIMO2019, mishraSumThroughputMaximization2019a} or sophisticated tag capabilities, such as decoding control signals from ambient illuminators~\cite{wangSpraySpectrumefficientAgile2024, bharadia_backfi_2015, zhang_freerider_2017, alim_backscatter_2017,yuanEnablingNativeWiFi2023}, coordinating with each other~\cite{choi_devices_2021}, or sensing the channels before transmitting~\cite{cao_distributed_2020,ma_design_2019}. While the prototype in \cite{alevizosMultistaticScatterRadio2018} addresses multi-static architecture and avoids requiring sophisticated capabilities at tags, it does require each tag to have a dedicated frequency allocation. This limits the number of tags that can operate simultaneously in the network. These assumptions make such multiple access strategies impractical for large-scale bi-static BNs with receiver-less tags. Moreover, none of these studies systematically evaluate the scalability of bi-static BNs.

	In bi-static BNs with receiver-less tags, only uncoordinated channel access can be supported, similar to the well-known ALOHA protocol. Two studies \cite{valentini_cross-layer_2020, koizumi_asynchronous_2023} have proposed using ALOHA-like uncoordinated channel access methods for bi-static BNs. However, both rely on a steady out-of-band link between illuminators and receivers, and \cite{koizumi_asynchronous_2023} further requires spatial separation of all tags to independently schedule illumination of each tag. Our work, by contrast, does not rely on coordination between the illuminator and receiver, significantly reducing deployment constraints. 
	
	Our work also introduces the concept of a ``collision zone parameter'', which quantifies the effect of packet collisions on packet corruption in an uncoordinated network. We present a practical methodology for estimating this parameter using a small-scale network prototype. This approach can be applied to any type of tag packet and serves as a valuable tool for evaluating network performance after tuning key network parameters (e.g., packet structure and error-correcting mechanisms). With this methodology, system designers can assess the impact of packet collisions and make rapid adjustments to network parameters, resulting in more efficient deployments.
	
	Finally, our work leverages an interesting theoretical insight: the network reliability with entirely uncoordinated channel access can be seen as depending solely on channel occupancy, independent of the number of devices. While this insight is theoretically well known, we use it to estimate the collision zone parameter on a smaller prototype to emulate large-scale networks. We believe this novel approach of experimentation offers a significant advantage in analyzing large-scale uncoordinated networks with real-world measurements – a contribution of independent interest to the field.

	\subsection{Contributions}
	Our contributions can be summarized as follows:
	\begin{itemize}
		\item We define a reliability metric called Tag Drop Rate (TDR), and introduce the \textit{collision zone parameter} that measures the link between packet collision and corruption in a network. 
		We derive the expression of TDR as a function of the network parameters and the collision zone parameter. 
		We show from the theoretical model that under certain assumptions, the reliability can be seen as only dependent on the channel occupancy and agnostic of the number of tags in the network. 
		\item We use this key insight to design a large-scale network using the measurements on a small-scale experimental testbed. Specifically, we measure the collision zone parameter with two small-scale prototypes of the bi-static BN. Our prototypes consist of wired and wireless setups with two USRP software-defined radios (SDRs) acting as the illuminator and the receiver, and 8 custom-built backscatter semi-passive tags. 
		\item We use the theoretical model and the collision zone parameter -- estimated from real-world measurements -- to determine the network parameters of a large-scale bi-static BN that satisfy the required reliability and bandwidth criteria. 
		\item Finally, we show that a BN with bi-static radio architecture and TDM as the multiple access scheme can reliably support a large-scale network with 1000s tags. 
	\end{itemize}
	\subsection{Outline}
	The remaining sections of this paper are organized as follows: In \secref{sec:system}, we describe the system model, introduce the network parameters, the reliability metric, and the design constraints. In \secref{sec:theory}, we introduce the collision zone parameter, derive the reliability of a bi-static BN, and discuss the insight that enables using small-scale prototypes to model large-scale networks. We then introduce the experimental testbeds used to estimate the collision zone parameter in \secref{sec:experiment}, followed by the evaluation results in \secref{sec:tuning}. We then evaluate the network performance of a large-scale bi-static BN using the calibrated theoretical model and simulations in \secref{sec:results}. Finally, we provide the concluding remarks in \secref{sec:discussion}.
	
	\section{System Model}\label{sec:system}
	\subsection{System parameters}
	We consider a bi-static system consisting of an illuminator (transmitter), a receiver, and $K$ passive backscatter tags. We expect $K$ to be in the orders of hundreds in a typical large-scale network. We assume the illuminator and the receiver are active for the duration of system operation. 
	
	\begin{figure}[tb]
		\centering
		\includegraphics[width=\columnwidth]{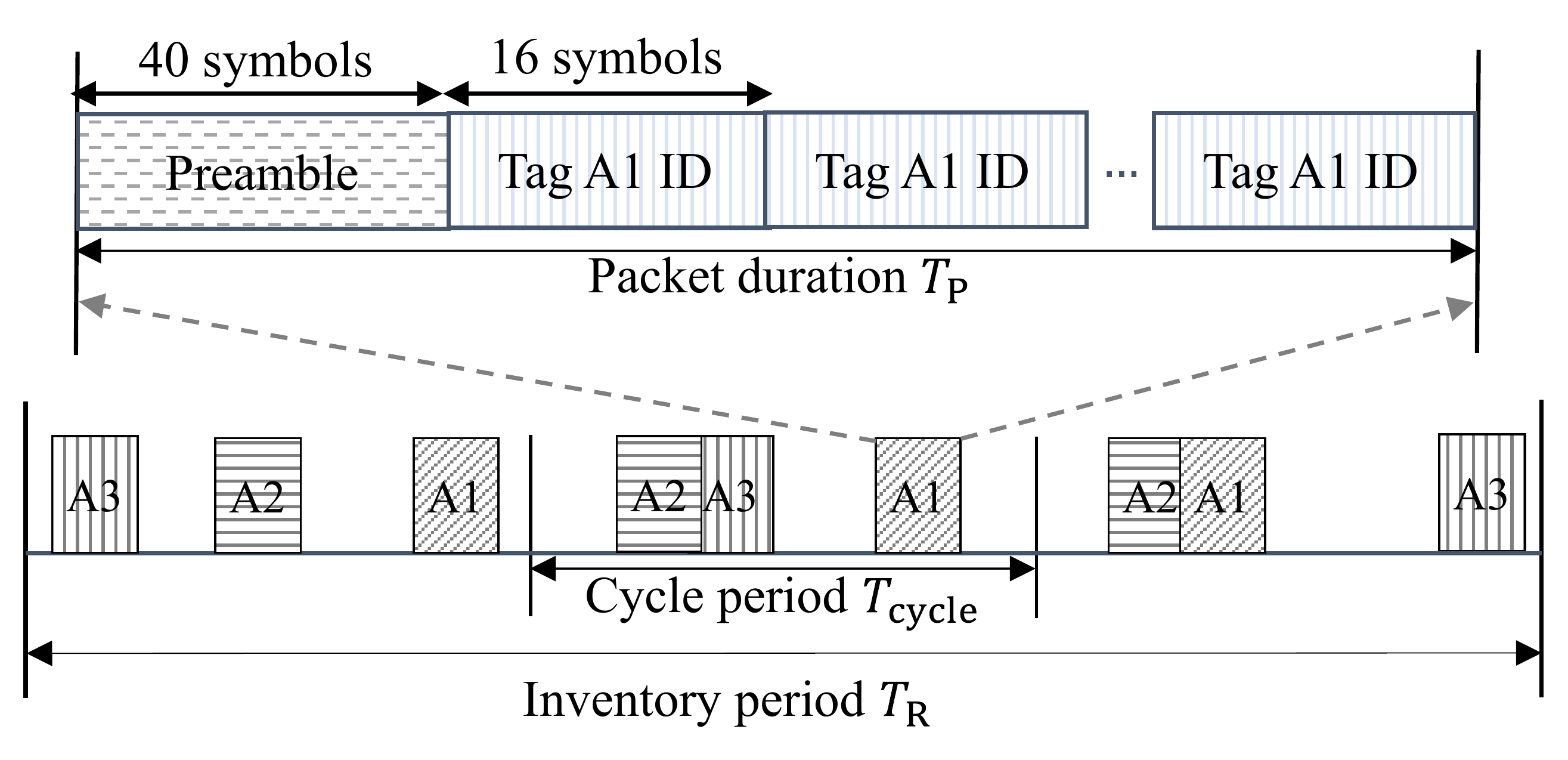}
		\caption{Packet structure, cycle period and inventory period for time multiplexed tags: The figure shows 3 tags, A1, A2, A3, operating with 3 cycles in one inventory period of duration $\TR$ ($\TR=3\Tcycle$).}
		\label{fig:packet}
	\end{figure}
	
	To minimize the coordination between the illuminator, receiver, and the tags, we consider un-slotted random access as the multiple access scheme for the tags. Specifically, each tag wakes up at a uniformly random time within a cycle of duration $\Tcycle$. The tag then backscatters one pre-recorded packet, called \textit{Tag Packet} (see \figref{fig:packet}). The tag packet contains a preamble of length 40 symbols and $\Nrep$  number of repetitions of tag ID of length 16 symbols. Assuming the symbol rate of $\Rs$, the packet duration can be defined as $\Tp=(40+16\Nrep)/\Rs$. For the simplicity of the modulating circuit for backscatter tags, we consider a square waveform to modulate the tag packet. Hence, to transmit the symbols at the rate $\Rs$, the tag uses the transmission bandwidth of $6\Rs$ as more than 96\% of energy of the square waveform concentrates in $6\Rs$ bandwidth.
	
	Note that tag packets that overlap in time create interference at the receiver. These collisions within a cycle fundamentally limit the MAC-layer performance of the network. The parameters, such as the number of tags $K$, packet duration $\Tp$ (characterized by $\Rs,\Nrep$), and cycle duration $\Tcycle$, can be optimized to control the collision in the network. In the following section, we define a metric of reliability of the network and discuss the constraints and associated trade-offs in tuning these parameters. 
	
	\subsection{Tag Drop Rate: A reliability metric}
	Let the \textit{inventory period} (or \textit{refresh period}), denoted by $\TR$, be defined as the duration within which all tags are required to transmit at least one tag packet. Then, we define TDR, denoted by $\Rdrop$, as a fraction of tags that could not transmit a single successful packet in one inventory period $\TR$ (i.e., the number of unsuccessful tags in an inventory period divided by the total number of tags). 
	
	Note that, for better reliability (i.e., lower TDR), we also allow a tag to transmit multiple packets in one inventory period. This means the cycle duration $\Tcycle\leq \TR$. For simplicity, we consider $\Tcycle$ such that there are $L$ cycles in each inventory period ($\TR=L\Tcycle$). Overall, the TDR requirement along with the inventory period $\TR$ provides the design requirement of the system. 
	
	\subsection{System specification and design trade-offs}
	In theory, provided enough bandwidth, an arbitrarily high data rate can be achieved, which leads to smaller packet duration, lower collisions, and lower TDR. In practice, however, wireless networks are constrained to operate in regulated frequency bands. Let $B$ MHz denote the maximum permissible bandwidth of operation. Then, from the network layer perspective, the bandwidth limitation defines the fundamental (and the only) constraint on the system. Specifically, the bandwidth limitation of $B$ MHz constrains the symbol rate $\Rs$ to be less than $B/6$ MBaud per second ($6\Rs\leq B$ MHz). 
	
	Finally, we formalize the system requirements. In this work, we design a wide-area BN (WABN) which satisfies the following specifications:
	\begin{enumerate}
		\item \textit{Scale} $K$: The network must support $K$ backscatter tags. In a typical industrial application, $K$ can be of the order of hundreds to thousands. 
		\item \textit{Reliability} $\delta,\TR$: The maximum $\Rdrop$ of the system with an inventory period $\TR$ should be less than $\delta$. 
	\end{enumerate}
	To design WABN with the specifications ($K,\delta,\TR$), we tune the following system parameters:
	\begin{enumerate}
		\item Packet duration $\Tp$ as a function of the number of ID repetitions $\Nrep$ and the symbol rate $\Rs<B/6$ MBaud per second.
		\item The number of cycles $L$ and cycle duration $\Tcycle$.
	\end{enumerate}
	
	These parameters introduce several trade-offs in the system. Firstly, a high number of tags $K$, a lower TDR requirement $\delta$, and a lower inventory period $\TR$ lead to a harder design requirement. Furthermore, for a given requirement of $(K,\delta,\TR)$, there exist interesting trade-offs in the system design: (1) error-correction for packet recovery using $\Nrep$: Higher $\Nrep$ can allow packet recovery in case of collisions, but it also increases the packet length which can result in a higher collision rate. (2) the number of attempts for successful packet transmission using $L$: Higher $L$ can reduce the TDR due to more attempts, but the smaller cycle duration can increase the number of collisions. Moreover, it is unclear whether the system should prioritize the error-correction-based packet recovery or simply increase the number of attempts by tags in each inventory period. Furthermore, the solution to these trade-offs depends on the symbol period $\Rs$, which is constrained by the low-energy requirement of the tags and spectrum regulations. To simplify these trade-offs, we develop a theoretical model that maps the reliability metric with the system parameters.

	\section{Theoretical Analysis}\label{sec:theory}
	Before delving into the theoretical analysis, we emphasize that the TDR, in a way, measures the corrupted tag packets. We, however, can only analyze the number of collisions in the network using probabilistic analytical frameworks. 
	Notably, not all collisions result in packet corruption (e.g., some collided tag packets can be decoded), and not all packet corruptions are due to collisions (e.g., they can be due to a noisy channel). 
	Therefore, we first analyze the impact of packet collisions on the corruption. 
	Thus, for the initial theoretical analysis, we consider a collision-limited or noise-free network such that all packet corruptions are due to collisions (BER = 0). 
	
	To quantify the effect of collisions on packet corruption, we define a collision zone parameter, denoted by $\alpha$, that characterizes an aggregate relationship between the collision and the corruption of packets in the network. 
	Specifically, the collision zone parameter $\alpha$ is defined as the average fraction of overlap permissible by two packets such that both packets are recoverable. For instance, if $\alpha=0.5$, then on average, two packets with more than 50\% of overlap are corrupted. We note that $\alpha=0.5$ is a macro-level statistic of the network and it does not imply that all packets are corrupted only if they have an overlap of more than 50\%. It is possible that a few packets with a smaller overlap may result in packet corruption (say, if the preamble is corrupted), and some with a larger overlap may not. 
	
	A key benefit of using such a macro-level statistic is that it can be estimated using experimental measurements and its estimate can allow modeling the network by abstracting underlying physical layer characteristics. 
	
	\subsection{Theoretical model of TDR}
	Consider a packet transmission by the $i$-th tag in a cycle. Let $t_i\sim \cc{U}(0,\Tcycle )$ be the time instance when the tag begins transmission of the packet, where $\cc{U}(a,b)$ denotes a continuous uniform distribution between $a$ and $b$. Given that all tags randomly and independently select the time for the packet transmission, all $t_i$’s are independent and identically distributed. 
	
	Let $E_j^i$ be the event that the packets transmitted by $i$-th tag and $j$-th tag are overlapped by more than $\alpha$ fraction. Hence,
	\begin{equation}
		E_j^i=\left\{(t_i,t_j ):|t_i-t_j |<(1-\alpha)\Tp \right\}.
	\end{equation}
	
	Let $p_j^i$ be the probability that the event $E_j^i$ occurs. Then, the probability that the packet of tag $j$ does not corrupt the packet of tag $i$ is given by
	\begin{equation}
		1-p_j^i=\left(1-(1-\alpha) \frac{\Tp}{\Tcycle} \right)^2.
	\end{equation}
	Let $\Dcycle=\Tp/\Tcycle$ denote the cycle occupancy of a tag and $\beta = 1-\alpha$. Then, the probability that no tag packet collides with the packet of tag $i$ in a cycle is given by
	
	\begin{equation} 
		P_i=\prod_{j\neq i} (1-p_j^i ) =(1-\beta \Dcycle )^{2(K-1)}.
	\end{equation}
	The probability that the packet of tag $i$ is corrupted in a cycle period is given by $(1-P_i )$. Subsequently, since all tags operate independently across cycle periods, the probability that the packet of tag $i$ is corrupted in all cycles of an inventory period is given by $(1-P_i )^L$. 
	
	We now derive the expression for the TDR $\Rdrop$. Let $X_i$ be a Bernoulli random variable such that $X_i=1$ if the packet of tag $i$ is not received by the end of an inventory period. Hence, $P(X_i=1)=(1-P_i )^L$. Using the definition of TDR, we can characterize $\Rdrop$ as follows:
	\begin{IEEEeqnarray}{rl}
		\Rdrop & =\mathbb{E}\left[\frac{1}{K} \sum_{i} X_i \right]=\frac{1}{K} \sum_i \mathbb{E}[X_i]\\
		& = \left(1-(1-\beta \Dcycle )^{2(K-1)} \right)^L.\label{eq:Rdrop}
	\end{IEEEeqnarray}
	Even with all system parameters, the TDR cannot be estimated without the collision zone parameter $\alpha$. Also, it is impractical to model the collision zone parameter $\alpha$ purely based on theory as it depends on the physical layer signal processing techniques (such as preamble recovery, channel estimation methods, etc.). Thus, in \secref{sec:experiment}, we develop a testbed to estimate the value of $\alpha$ and, consequently, experimentally evaluate the performance of a bi-static BN.
	
	Note that the random-access technique in one cycle is indeed equivalent to an unslotted-ALOHA technique without re-transmission. Our analysis of the technique, however, uses a different approach. Specifically, the analysis of the conventional ALOHA scheme focuses on optimizing the capacity of the channel. That requires a channel-centric approach to model the periods without any packet transmission or with only one packet transmission. We instead focus on the reliability of the network. Therefore, we use a tag-centric approach and estimate the number of times a packet from the tag is received without the collision.

	\begin{figure}
		\centering
		\includegraphics[width=.75\columnwidth]{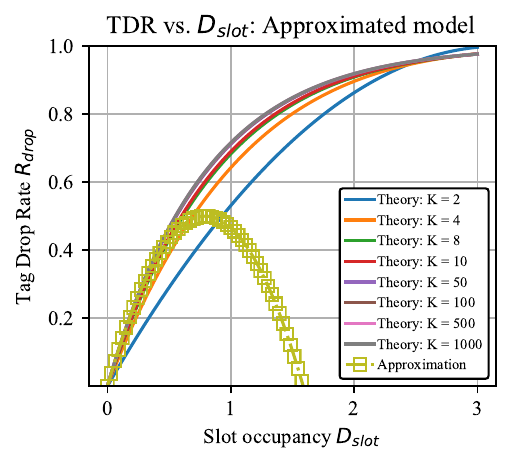}
		\caption{TDR vs. $\Dslot$ when $L = 1$ and $\hat{\alpha}=0.37$: The approximation is valid for small values of $\Dslot$. In this regime, TDR can be considered as a function $\Dslot$ and not the number of tags $K$.}
		\label{fig:approxModel}
	\end{figure}
	
	\subsection{Perspectives on TDR – System design and experiments}
	It is infeasible to emulate a large-scale BN with hundreds of backscatter tags on a testbed with a limited number of tags. Thus, to allow emulating a large WABN network on our testbed, assuming $K>>1$, we can re-write \eqref{eq:Rdrop} using Taylor series expansion as follows:
	\begin{equation}\label{eq:RdropApprox1}
		\Rdrop \approx \left(2\beta K\Dcycle-2(\beta K\Dcycle )^2 \right)^L  . 
	\end{equation}
	
	Note that the approximation deviates less than 5\% from the full equation when $\beta K \Dcycle<0.35$. Let $\Dinv=\Tp/\TR=\Dcycle/L$ denote the inventory occupancy of a tag. Furthermore, let $\Dslot=K\Dinv$ denote the slot occupancy, the average fraction of ``slots'' occupied within an inventory period. Hence, with the observation $K\Dcycle=L\Dslot$, we can re-write \eqref{eq:RdropApprox1} as
	\begin{equation}\label{eq:RdropApprox2}
		\Rdrop\approx\left[2\beta L\Dslot-2(\beta L \Dslot )^2 \right]^L.  
	\end{equation}
	Although, the derived theoretical model in \eqref{eq:Rdrop} shows the TDR depends on both $\Dslot$ and $K$, \eqref{eq:RdropApprox2} suggests that the TDR can be considered as a function of $\Dslot$ and independent of $K$. In \figref{fig:approxModel}, we plot TDR as a function of the slot occupancy $\Dslot$ and the number of tags $K$ with $L = 1$. We note from the figure that \eqref{eq:RdropApprox2} is valid for TDR $\Rdrop<0.4$, which is well above the typical requirements of TDR $\Rdrop<0.05$. Thus, for all practical purposes, TDR can be considered as a function of $\Dslot$ without needing the tag count $K$ in the network.
	
	This leads to a key insight: While \eqref{eq:RdropApprox1} models TDR as a function of the number of tags $K$ and the cycle occupancy of the tag $\Dcycle$ which directly relates to physical quantities, such as the power consumption of tags, \eqref{eq:RdropApprox2} models TDR invariant of the number of tags $K$. For instance, it suggests that the optimal number of cycles $L$ that minimizes TDR only depends on the slot occupancy $\Dslot$. Thus, \eqref{eq:RdropApprox2} enables modeling and validating a large-scale BN with hundreds of backscatter tags using a small testbed with only a few backscatter tags. We use this key idea in deriving the results of our testbeds for large-scale BNs in \secref{sec:tuning}.

	Finally, \eqref{eq:RdropApprox2} also provides an elementary approach for designing a network with specifications ($K,\delta,\TR$): (a) find the feasible $L$ and $\Dslot$ for a given $\delta$ using \eqref{eq:RdropApprox2}; (b) find the required $\Tp < \TR \Dslot/K$; (c) ensure $\Rs\leq B/6$ MBaud per second. In \secref{sec:results}, we provide an extension of this procedure that also accounts for the impact of BER.

	\section{Experimental Testbeds}\label{sec:experiment}
	In this section, we describe two testbeds used for evaluating the network performance and estimating the collision zone parameter. Since the collision zone parameter is meant to model the effect of the collisions on the corruption, we design the two testbeds to maintain nearly zero BER. In the following, we describe our setup, methodology, and two testbeds.
	
	\begin{figure*}[tb]
		\centering
		\includegraphics[width=.95\textwidth]{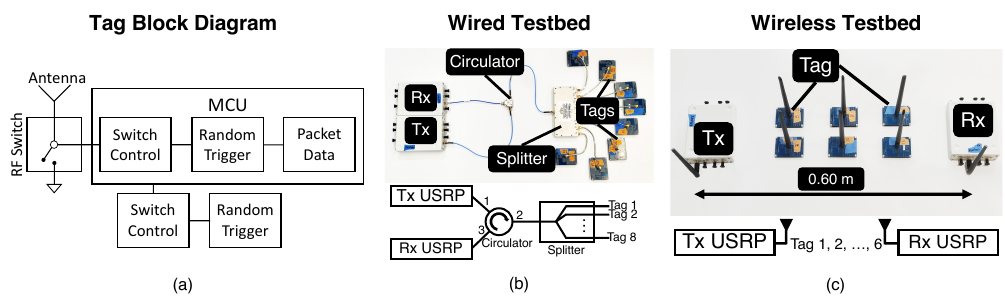}
		\caption{Experimental setup. (a) Simplified block diagram of a tag. (b) Wired testbed with 8 physical tags multiplexed by an RF splitter. A circulator connects the tags, transmitter USRP, and receiver USRP. (c) Wireless testbed with 6 physical tags placed between the transmitter USRP and receiver USRP, which are 0.6 m separated from each other. }
		\label{fig:experimentSetup}
	\end{figure*}
	
	\subsection{Hardware setup}
	We use two USRP SDR B210s as our illuminator and receiver, running on their internal clocks. We use Intel NUC8i7HVK1-based desktop running Ubuntu 18.04 and GNU Radio to control the USRPs. We use up to 8 custom-designed semi-passive backscatter tags powered with CR2032 lithium cell batteries. The operating frequency of the system is set to be 904 MHz. The raw signals are collected into binary files and processed. 
	
	\subsection{Backscatter tag}
	A simplified block diagram of our prototyped Gen 1 tags is depicted in \figref{fig:experimentSetup}(a). We use a single pole single throw RF switch to alternate the termination load of the antenna connected to the tag. Thus, the incident carrier wave on the antenna is backscattered with a different phase and amplitude corresponding to the antenna termination load. A microcontroller unit (MCU) is responsible for generating the tag data and for controlling the RF switch state. The tag data is generated with a random interval using a pseudorandom generator programmed in the firmware of the MCU. Finally, for powering the tag, a CR2032 battery is used with its voltage regulated to 1.8V using a low drop-out regulator. 
	
	\subsection{Measurement procedure}
	We evaluate the TDR $\Rdrop$, the metric for the system performance, as a function of the number of cycles $L$ and the slot occupancy $\Dslot$. 
	
	For the experimental evaluation, we first program all the tags to transmit with a packet duration $\Tp=1$ ms by fixing $\Rs=200$ kBaud per second and $\Nrep=10$, resulting in a packet with $200$ symbols. Then for each value of $K$ and $\Tcycle$, we turn on the illuminator and run the receiver to collect the raw I/Q samples generated by the backscatter tags. We define one measurement duration such that around $140$ packets are collected from each tag. The raw samples are then post-processed for packet detection and decoded to identify the ID of the tag that transmitted them. Notably, to test the collision-limited BN, we do not use any error correction procedure. Therefore, packets with even a one-bit error are discarded, assuming packet corruption. Finally, we list all the received along with their received timestamps and the tag IDs. 
	
	To measure the TDR, we first group all packets received in non-overlapping windows of duration $L\Tcycle$, where $L$ is the number of cycles. Each window of duration $L\Tcycle$ defines an inventory period. We count the number of tags which does not appear in an inventory period. The average of this number across all inventory periods gives the estimated TDR for given values of $L$ and $\Dslot$. We repeat the above process for different values of $L$ and $\Dcycle$.

	\subsection{Wired testbed}
	The wired testbed is built to completely exclude air as a medium. In this case, the system is expected to have zero BER and should perform very close to the theoretical analysis. 
	
	Our wired testbed is shown in \figref{fig:experimentSetup}(b). Specifically, we use an RF circulator to connect the illuminator USRP, the receiver USRP, and the tags. We use up to $8$ physical tags in this experiment, and they are further multiplexed by an $8-1$ RF power splitter. The sampling rate of USRPs is set to $2$ Msamples per second. For this wired testbed, we set the gain of the illuminator and receiver USRPs to $30$ dB. 
	
	\subsection{Wireless testbed}
	Our wireless testbed is shown in  \figref{fig:experimentSetup}(c), where the illuminator and receiver USRPs are separated by $0.6$ meters to ensure nearly zero BER. We put up to $6$ tags between the illuminator and receivers, and they are multiplexed over the air. For this wireless testbed, we set the gain of the illuminator and the receiver to $25$ dB and $70$ dB, respectively.
	\begin{figure*}[tb]
		\centering
		\includegraphics[width=.8\textwidth]{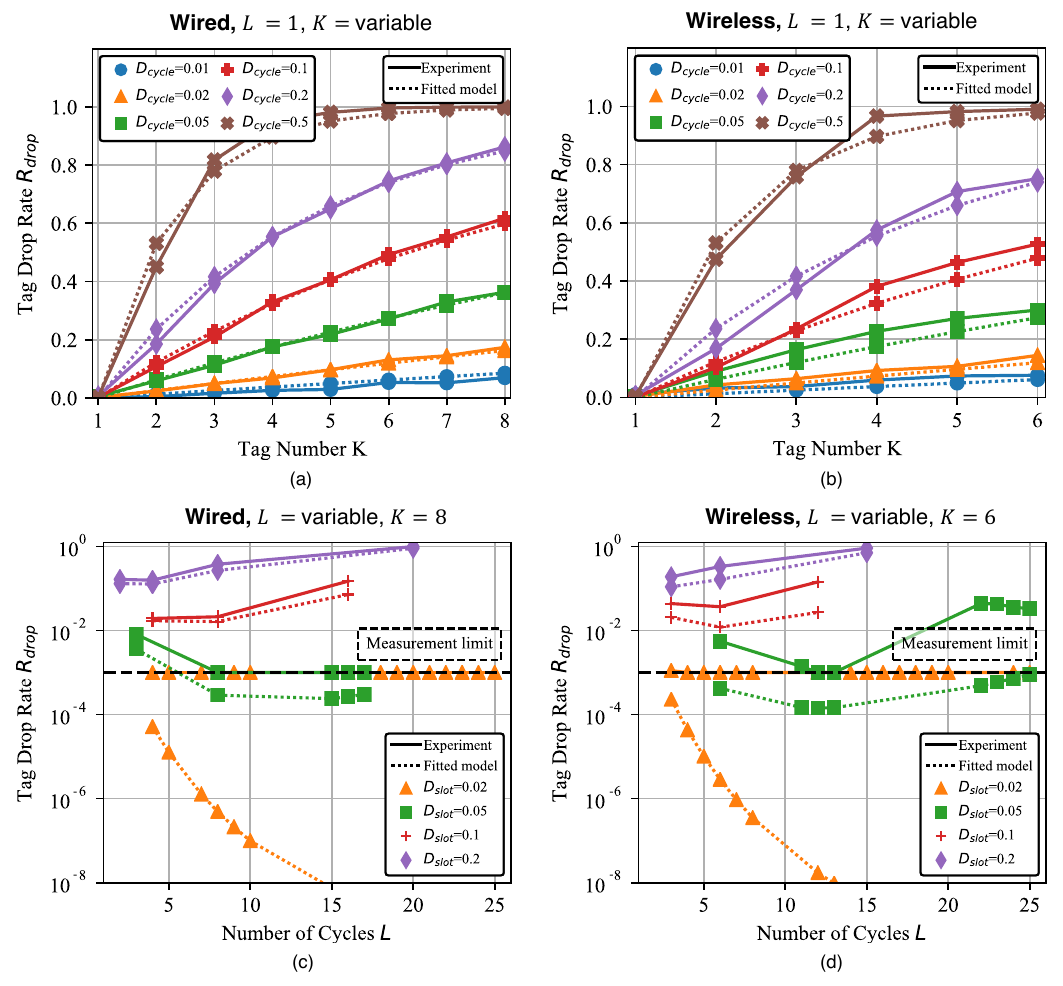}
		\caption{TDR: Plotted versus tag number $K$ for fixed $L=1$ with a (a) wired and (b) wireless testbed. Plotted versus number of cycles $L$ for fixed $K$ in a (c) wired, $K = 8$ and (b) wireless, $K = 6$ testbed. Solid lines show experimental results, dotted lines show theoretical values for $\hat{\alpha}=0.37$. Note that the precision of measured TDR is limited to $10^{-3}$ due to collection of 140 packets per tag. The deviation of experimental results from the theoretical model on the wireless testbed is due to the channel noise as the theoretical model assumes a noiseless channel. }
		\label{fig:experiment_results}
		\vspace{-2pt}
	\end{figure*}
	\section{Evaluation Results}\label{sec:tuning}
	In this section, we first map the theoretical model to the experimental results and show that the proposed theoretical model can measure TDR by estimating the collision zone parameter $\alpha$. We then further discuss the trends of the theoretical model as corroborated by the experimental results.
	
	\subsection{Estimating the collision zone parameter $\alpha$}\label{sec:estimateAlpha}
	Recall that the collision zone parameter $\alpha$ defines the amount of collision between packets that leads to corruption. In this section, we describe the procedure to estimate the value of $\alpha$ using the experimentally measured TDR $\Rdrop$. 
	
	Since the wired setup is a collision-limited system, not a noise-limited one, the theoretical model which assumes a collision-limited system, can be mapped onto experimental observations. Accordingly, we use the data collected from the wired setup to estimate the collision zone parameter $\alpha$ using the theoretical model derived in \secref{sec:theory}. Specifically, we use \eqref{eq:Rdrop} to calculate the theoretical TDR and compare this theoretical value with the experimental TDR data to get the fitting error – we use Root Mean Squared Error (RMSE) as our evaluation metric. By numerically testing values of collision zone parameter $\alpha$, we get the minimum RMSE at value $\hat{\alpha}=0.37$. Therefore, we can model a collision of packets resulting in corruption if the packets are overlapped by more than 37\%.
	
	Again, we emphasize that the collision zone parameter is an aggregate metric, i.e., some packets with a smaller overlap may result in packet corruption (say, if the preamble is corrupted), and some packets with a larger overlap may not. In addition, the value of the collision zone parameter depends on the packet structure and receiver design. For instance, by using successive interference cancellation at the receiver~\cite{sen_successive_2010} or the coding schemes for backscatter applications~\cite{rezaei_coding_2023}, the chance of packet recovery after the collisions can be increased, which can increase the estimated collision zone parameter $\hat{\alpha}$. 
	
	In Figs. \ref{fig:experiment_results}(a) and \ref{fig:experiment_results}(c), we plot TDR evaluated using the measurements from the wired testbed and the theoretical model with the estimated $\hat{\alpha}$. They confirm that evaluated TDR closely follows the theoretical model. In Figs. \ref{fig:experiment_results}(b) and \ref{fig:experiment_results}(d), we plot TDR evaluated using the data from the wireless testbed and compare it with the theoretical model. In this case, we observe the deviation of experimental results from the theoretical model which can be attributed to the slight noise in the wireless channel. Since our theoretical analysis assumed a zero-noise channel, the experimental TDR is higher than the theoretical estimate. % We further analyze the impact of the noise on the network in \secref{sec:results}. 

	\subsection{Tag drop rate with $L=1$}
	We first consider the TDR when $L=1$. In this case, each tag only makes a single attempt to transmit a packet during each inventory period. The tag is considered dropped if the receiver cannot receive and decode this packet. In Figs. \ref{fig:experiment_results}(a) and \ref{fig:experiment_results}(b), we show the TDR measured on the wired and wireless testbed for different $K$ and $\Dcycle$. Unsurprisingly, increasing the number of tags $K$ or the cycle occupancy $\Dcycle$, the reliability degrades due to higher congestion in the network.
	
	\begin{figure}[tb]
		\centering
		\includegraphics[width=.75\columnwidth]{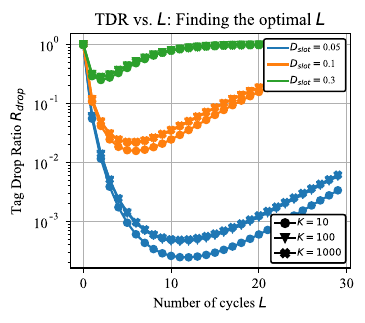}
		\caption{Theoretical TDR vs. the number of cycles $L$: The optimal number of cycles primarily depend on $\Dslot$ as observed from \eqref{eq:RdropApprox2}.}
		\label{fig:optimalL}
	\end{figure}
	\subsection{Tag drop rate with $L>1$}
	In Figs. \ref{fig:experiment_results}(c) and \ref{fig:experiment_results}(d), we plot TDR (in log-scale) with $L>1$, where each tag makes multiple attempts per inventory period. We can observe that a higher slot occupancy $\Dslot$ degrades the performance of the network in both wired and wireless settings. We can also see a trade-off: the higher value of $L$ results in more congestion in each cycle; thus, more tag packets are corrupted in each cycle, however, a small value of $L$ results in fewer attempts by tags in an inventory period, thus, impacting the TDR. We can also observe that there exists a non-trivial value of the optimal number of cycles $L$ which depends on $\Dslot$. We use simulation to further analyze the impact of $L$ on the TDR and identify the optimal number of cycles $L$ in \secref{sec:results:ImpactOfL}.
	
	\begin{remark}
		We emphasize that while we use semi-passive tags to estimate the collision zone parameter, the scalability analysis presented in this work can also be applied to passive tags. Specifically, the MAC protocol, the estimated collision zone parameter, and the resulting numerical results are independent of whether the tag is semi-passive or passive. The MAC protocol requires the tag to be active for a duration of $\Tp$ in every inventory period $\TR$, which can be either powered by the battery or the carrier/illuminator signal for semi-passive or passive tags, respectively. Packet collisions and their impact on packet corruption are also independent of the tag's power source. Therefore, the estimated collision zone parameter is not affected by the power source of the tag. Consequently, the scalability analysis presented in this work can be applied to the settings with passive tags.
	\end{remark}
	
	\begin{figure}[tb]
		\centering
		\includegraphics[width=.75\columnwidth]{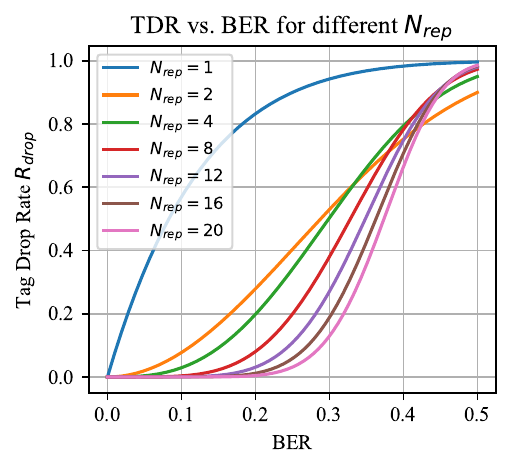}
		\caption{TDR vs. BER of the channel in the noise-limited network $(L=1,K=1)$: Increasing $\Nrep$ improves the tolerance for BER for a fix TDR.}
		\label{fig:noise-limited}
	\end{figure}
	
	\section{Numerical Analysis}\label{sec:results}
	In this section, we extend our analysis to a more general setting with noise and collision-based packet corruption. We first discuss the impact of the number of cycles $L$ on the TDR in the collision-limited setting. We then analyze the TDR with varied BER and packet outages in a noise-limited setting. Based on that, we describe the procedure for designing a system with specific requirements and derive physical layer requirements that can be used for network deployment. 
	
	\subsection{Collision-limited network: Impact of the number of cycles}\label{sec:results:ImpactOfL}
	In \figref{fig:optimalL}, we plot the theoretical TDR with $\hat{\alpha}=0.37$ as a function of the number of cycles $L$ for different values of $\Dslot$ and $K$. Note that the optimal number of cycles $L$ primarily depends on $\Dslot$, and it reduces with increasing value of $\Dslot$. This observation can serve as a key design step for a large-scale BN. Specifically for the reliability requirement of $\Rdrop<\delta$ with $K$ tags, we can find optimal $L,\Dslot$. Then, we can use the feasible values of $L,\Dslot$ to find the system parameters such as packet length $\Tp$ and symbol rate $\Rs$. 
	
	\subsection{Noise-limited network: Impact of the repetitions}
	
	Before delving into full-scale system simulation, we first analyze the impact of the channel noise, i.e., BER, on the TDR in a collision-free or noise-limited network. For that, we consider a collision-free network with only one tag. Since there are no collisions in the network, the TDR is only determined by BER and can therefore be reduced exponentially by increasing the number of cycles $L$. For clarity of the discussion, we consider $L=1$. In \figref{fig:noise-limited}, we plot TDR with respect to the BER of the channel for different numbers of repetitions of the tag ID in the packet which characterizes the performance of repetition codes. Unsurprisingly, we observe that increasing $\Nrep$ allows a higher tolerance for BER for a given requirement of TDR. 
	
	Upon understanding the impact of two parameters, $\Nrep$ and $L$ in collision-limited and noise-limited settings, respectively, we now simulate a realistic network by considering both, the collision in the network and the noise of the channel. 
	
	\begin{figure}[tb]
		\centering
		\includegraphics[width=.85\columnwidth]{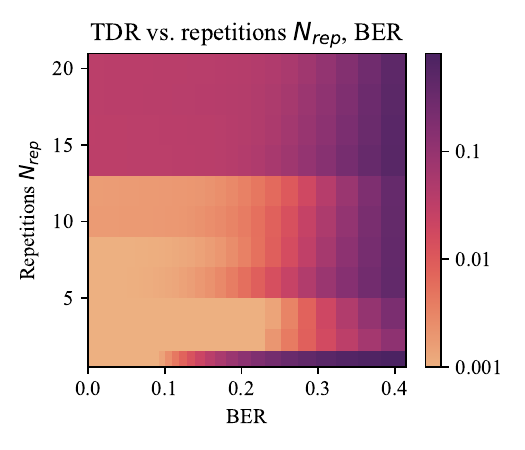}
		\caption{TDR vs. Number of tag ID repetitions in the packet and BER of the channel in a general setting: The bright area with TDR $\Rdrop\leq 0.1$\% shows feasible region. The figure also emphasizes the impact of the collision. Unlike the noise-only setting, increasing $\Nrep$ can increase TDR due to higher collisions.}
		\label{fig:finalResult}
	\end{figure}
	
	\subsection{Simulation setup for system design}
	We design a BN for an asset tracking application that consists of 1000 tags and requires, on average, 99.9\% of all tags must be identified and located in each second. From the network layer perspective, this requires TDR $\Rdrop <0.001$ with an inventory period of $\TR=1$ s. Furthermore, we assume the symbol rate $\Rs=2$ MBaud per second which keeps the operational bandwidth of $B=12$ MHz within the ISM band.
	
	We consider the number of repetitions of Tag ID ($\Nrep$) and the number of cycles $L$ as the tuning parameters. We assume the channel noise introduces BER of $\gamma$. 
	
	Following the system model, in the simulation, we assume that all tags wake up at a uniformly random time in each cycle period $\Tcycle=1/L$ s, backscatter their respective tag packets, and sleep. Furthermore, if two packets are overlapped by more than $\hat{\alpha}=0.37$ fraction (derived from experimental observation), both packets are considered corrupted. In addition to that, we model the channel noise by randomly flipping the non-overlapping data bits with probability $\gamma$. We then calculate the probability of missed identification of a tag packet after the error correction procedure. With this probability, called packet error rate (PER), we discard non-corrupted packets, emulating unrecoverable packets due to the noise in a real network. Finally, we estimate TDR $\Rdrop$ in the system.

	\subsection{System design: Tuning $\Nrep$ and $L$}

	We now tune the system parameters $\Nrep$ and $L$ to achieve TDR $\Rdrop<\delta=0.001$ for different values of BER $\gamma$. Specifically, for each BER, $\Nrep$, we estimate TDR of the network with $L=\{1,\ldots,15\}$. In \figref{fig:finalResult}, we plot the minimum achievable TDR (at the optimal $L$) for each value of $\Nrep$ and BER $\gamma$. Thus, any point $(\Nrep,\gamma)$ in the plot that achieves TDR $\Rdrop <0.001$ indicates there exists $L$ such that with $L$ cycles in an inventory period, $\Nrep$ number of repetitions in the packet and the BER of $\gamma$, the system requirement can be satisfied. Note that, given the values of $\Nrep$ and $L$, all system parameters defined in \secref{sec:system} can be specified. We summarize a subset of feasible values of $\Nrep,L$ in \tblref{tab:table1}, along with other system parameters and maximum permissible BER.
	
	\begin{table}[tb]
		\centering
		\caption{Network parameters that satisfy $\Rdrop<0.1$\% with $\TR=1$ s.}
		\resizebox{\linewidth}{!}{%
			\begin{tabular}{|c|c|c|c|c|} 
				\hline
				\begin{tabular}[c]{@{}c@{}}\textbf{Cycles} \\$L$\end{tabular} & \begin{tabular}[c]{@{}c@{}}\textbf{Cycle }\\\textbf{duration }\\$\Tcycle$\textbf{(s)}\end{tabular} & \begin{tabular}[c]{@{}c@{}}\textbf{Number of}\\\textbf{ repetitions}\\ $\Nrep$ \end{tabular}& \begin{tabular}[c]{@{}c@{}}\textbf{Packet}\\\textbf{duration}\\ $\Tp$\textbf{~(ms)} \end{tabular} & \begin{tabular}[c]{@{}c@{}}\textbf{Maximum}\\\textbf{~tolerable }\\\textbf{BER}\end{tabular} \\ 
				\hline
				2 & 0.500 & 4 & 0.052 & 0.0746 \\ 
				\hline
				4 & 0.250 & 4 & 0.052 & 0.1585 \\ 
				\hline
				6 & 0.167 & 4 & 0.052 & 0.2037 \\ 
				\hline
				8 & 0.125 & 4 & 0.052 & 0.2215 \\ 
				\hline
				10 & 0.100 & 4 & 0.052 & 0.2215 \\ 
				\hline
				11 & 0.091 & 4 & 0.052 & 0.2215 \\ 
				\hline
				\textbf{12} & \textbf{0.083} & \textbf{4} & \textbf{0.052} & \textbf{0.2215} \\ 
				\hline
				13 & 0.077 & 4 & 0.052 & 0.2037 \\ 
				\hline
				14 & 0.071 & 2 & 0.036 & 0.1723 \\ 
				\hline
				15 & 0.067 & 2 & 0.036 & 0.1458 \\
				\hline
			\end{tabular}
		}
		\label{tab:table1}
	\end{table}
	
	It is important to emphasize that out of all feasible values of $(\Nrep,L)$, the pair that achieves $\Rdrop<0.001$ with the highest permissible BER is preferred. A higher (i.e., lenient) BER requirement allows network deployment with a lower noise floor, leading to a larger coverage area and fewer illuminators and receivers. From the \tblref{tab:table1}, we conclude that a deployment that maintains BER below 22\% across the required area can fulfill the requirement of $\Rdrop<0.001$ with 1000 tags and 1 second of inventory period. Furthermore, the bold row in \tblref{tab:table1} shows the associated network parameters.

	\section{Conclusion and Future Work}\label{sec:discussion}
	In this work, we showed the feasibility of a bi-static BN for a large-scale wide-area IoT application. Specifically, we analyzed the performance of the network in terms of reliability. We derived a theoretical model of TDR as a function of network parameters and tuned it using experimental measurements. Based on the theoretical and experimental analysis, we provided a systematic framework for network design based on the specifications. Overall, we find that even with a modest physical layer requirement of BER below 22\% and the bandwidth of 12 MHz, the bi-static architecture can support a network of 1000 tags with TDR $\Rdrop < 0.001$ and the inventory period of $1$ s. 
	
	To fully deploy the network, we believe there are still several open questions that need to be answered. We summarize the key questions and the future directions below. 
	
	\subsection{Physical layer design of the WABN}
	The numerical analysis provides the requirement of BER. Thus, the immediate next step is the deployment design of the network. 
	The deployment requires considering the physical layer aspects of the network by accounting for the link budget of backscatter tags, their radiation patterns, multi-path effects in the environment, the energy source of the tags, and the illuminator interference on the receiver. Furthermore, the deployment design can incorporate the extension to a multi-static setting - using different frequencies on different illuminators to frequency multiplex the tags that are excited by the respective illuminators albeit with only one receiver. 
	Deployment planning for a specific real-world use-case will also offer an insight into the network life which measures the duration for which the tag can stay on with a given power source. 
	Collectively, the coverage of the deployment plan, hardware costs, and the network life will provide an ultimate physical layer feasibility test for the bi-static BNs in real-world use-cases. 
	
	\subsection{Efficient error correction coding}
	In the prototype of our tags, we have considered repeating the tag ID for data recovery in the case of collisions and noisy channels. The repetition codes, however, are known to have a low code rate, i.e., higher overhead. A more sophisticated error correction technique, with efficient coding rates, can potentially reduce the overhead of error corrections for the same tolerance to BER. Therefore, for the same reliability requirement, the system may support the higher tolerance of BER for the same packet length (and collision rate). Furthermore, recent research in error correction codes for collision-recovery is promising to further improve the network performance~\cite{rezaei_coding_2023}.
	
	\subsection{Improved packet recovery}
	As discussed in \secref{sec:estimateAlpha}, there are several ways to further increase the capacity of the network by employing the sophisticated packet recovery algorithm after the collision. For instance, successive interference cancellation~\cite{sen_successive_2010}, coding for improved tag recovery~\cite{mahdavifar_coding_2015}, can further be explored.
	
	\subsection{Spatial multiplexing at the receiver}
	Using a directional, array-based receiver can limit the number of simultaneous tag reads by limiting the region for capturing the backscattered signals (as explored in \cite{deavours_application_2009}), which reduces tag collisions, the risk of packet corruption and tag drops. This approach, however, introduces an additional constraint on the design. For instance, if the receiver uses a static round-robin schedule to read the tags from different regions, the tags within each region must respond within a shorter timeframe (i.e., smaller $\Tcycle$). Investigating the influence of the receiver's schedule for spatial multiplexing on TDR as a function of the spatial distribution of the tags for various use cases can be a valuable subsequent study.
	
	\bibliographystyle{IEEEtran}
	\bibliography{main}

\begin{thebibliography}{10}
\providecommand{\url}[1]{#1}
\csname url@samestyle\endcsname
\providecommand{\newblock}{\relax}
\providecommand{\bibinfo}[2]{#2}
\providecommand{\BIBentrySTDinterwordspacing}{\spaceskip=0pt\relax}
\providecommand{\BIBentryALTinterwordstretchfactor}{4}
\providecommand{\BIBentryALTinterwordspacing}{\spaceskip=\fontdimen2\font plus
\BIBentryALTinterwordstretchfactor\fontdimen3\font minus
  \fontdimen4\font\relax}
\providecommand{\BIBforeignlanguage}[2]{{%
\expandafter\ifx\csname l@#1\endcsname\relax
\typeout{** WARNING: IEEEtran.bst: No hyphenation pattern has been}%
\typeout{** loaded for the language `#1'. Using the pattern for}%
\typeout{** the default language instead.}%
\else
\language=\csname l@#1\endcsname
\fi
#2}}
\providecommand{\BIBdecl}{\relax}
\BIBdecl

\bibitem{RFIDConf24MAC}
K.~Patel, J.~Zhang, I.~Kiminos, L.~Kampianakis, M.~Eggleston, and J.~Du,
  ``Poster: Evaluating scalability of a large-scale bi-static backscatter
  network,'' \emph{2024 IEEE International Conference on RFID}, June 2024.

\bibitem{vannucci_software-defined_2008}
G.~Vannucci, A.~Bletsas, and D.~Leigh, ``A software-defined radio system for
  backscatter sensor networks,'' \emph{IEEE Transactions on Wireless
  Communications}, vol.~7, no.~6, pp. 2170--2179, Jun. 2008.

\bibitem{van_huynh_ambient_2018}
N.~Van~Huynh, D.~T. Hoang, X.~Lu, D.~Niyato, P.~Wang, and D.~I. Kim, ``Ambient
  backscatter communications: A contemporary survey,'' \emph{IEEE
  Communications Surveys \& Tutorials}, vol.~20, no.~4, pp. 2889--2922, 2018.

\bibitem{kampianakis_wireless_2014}
E.~Kampianakis, J.~Kimionis, K.~Tountas, C.~Konstantopoulos, E.~Koutroulis, and
  A.~Bletsas, ``Wireless environmental sensor networking with analog scatter
  radio and timer principles,'' \emph{IEEE Sensors Journal}, vol.~14, no.~10,
  pp. 3365--3376, Oct. 2014.

\bibitem{jiangBackscatterCommunicationMeets2023}
T.~Jiang, Y.~Zhang, W.~Ma, M.~Peng, Y.~Peng, M.~Feng, and G.~Liu, ``Backscatter
  communication meets practical battery-free internet of things: A survey and
  outlook,'' vol.~25, no.~3, pp. 2021--2051.

\bibitem{chawla_overview_2007}
V.~Chawla and D.~S. Ha, ``An overview of passive {RFID},'' \emph{IEEE
  Communications Magazine}, vol.~45, no.~9, pp. 11--17, Sep. 2007.

\bibitem{daniel_dobkin_rf_2012}
{Daniel Dobkin}, \emph{The {RF} in {RFID}}, 2nd~ed., ser. {UHF} {RFID} in
  {Practice}, Nov. 2012.

\bibitem{wang_link_2008}
H.~G. Wang, C.~X. Pei, and C.~H. Zhu, ``A link analysis for passive {UHF}
  {RFID} system in {LOS} indoor environment,'' in \emph{2008 4th
  {International} {Conference} on {Wireless} {Communications}, {Networking} and
  {Mobile} {Computing}}, Oct. 2008, pp. 1--7.

\bibitem{griffin_complete_2009}
J.~D. Griffin and G.~D. Durgin, ``Complete link budgets for backscatter-radio
  and {RFID} systems,'' \emph{IEEE Antennas and Propagation Magazine}, vol.~51,
  no.~2, pp. 11--25, Apr. 2009.

\bibitem{kimionis_increased_2014}
J.~Kimionis, A.~Bletsas, and J.~N. Sahalos, ``Increased range bistatic scatter
  radio,'' \emph{IEEE Transactions on Communications}, vol.~62, no.~3, pp.
  1091--1104, Mar. 2014.

\bibitem{vougioukas_could_2016}
G.~Vougioukas, S.-N. Daskalakis, and A.~Bletsas, ``Could battery-less scatter
  radio tags achieve 270-meter range?'' in \emph{2016 {IEEE} {Wireless} {Power}
  {Transfer} {Conference} ({WPTC})}, May 2016, pp. 1--3.

\bibitem{global_epc_epc_2008}
\BIBentryALTinterwordspacing
{Global, EPC}, ``{EPC} radio-frequency identity protocols class-1 generation-2
  {UHF} {RFID} protocol for communications at 860 {MHz}--960 {MHz},''
  \emph{GS1}, 2008. [Online]. Available:
  \url{https://www.gs1.org/sites/default/files/docs/epc/gs1-epc-gen2v2-uhf-airinterface_i21_r_2018-09-04.pdf}
\BIBentrySTDinterwordspacing

\bibitem{alevizosMultistaticScatterRadio2018}
P.~N. Alevizos, K.~Tountas, and A.~Bletsas, ``Multistatic {{Scatter Radio
  Sensor Networks}} for {{Extended Coverage}},'' \emph{IEEE Transactions on
  Wireless Communications}, vol.~17, no.~7, pp. 4522--4535, Jul. 2018.

\bibitem{buttAmbientIoTMissing2024}
M.~M. Butt, N.~R. Mangalvedhe, N.~K. Pratas, J.~Harrebek, J.~Kimionis,
  M.~Tayyab, O.-E. Barbu, R.~Ratasuk, and B.~Vejlgaard, ``Ambient {IoT}: A
  missing link in {3GPP IoT} devices landscape,'' vol.~7, no.~2, pp. 85--92.

\bibitem{alevizos_batteryless_2023}
P.~N. Alevizos, G.~Vougioukas, and A.~Bletsas, ``Batteryless backscatter sensor
  networks--part {I}: Challenges with (really) simple tags,'' \emph{IEEE
  Communications Letters}, vol.~27, no.~3, pp. 763--767, Mar. 2023.

\bibitem{zhang_hitchhike_2016}
P.~Zhang, D.~Bharadia, K.~Joshi, and S.~Katti, ``{HitchHike}: {Practical}
  backscatter using commodity {WiFi},'' in \emph{Proceedings of the 14th {ACM}
  {Conference} on {Embedded} {Network} {Sensor} {Systems}}, ser. {SenSys}
  '16.\hskip 1em plus 0.5em minus 0.4em\relax New York, NY, USA: Association
  for Computing Machinery, Nov. 2016, pp. 259--271.

\bibitem{zhang_freerider_2017}
P.~Zhang, C.~Josephson, D.~Bharadia, and S.~Katti, ``{FreeRider}: {Backscatter}
  communication using commodity radios,'' in \emph{Proceedings of the 13th
  {International} {Conference} on emerging {Networking} {EXperiments} and
  {Technologies}}, ser. {CoNEXT} '17.\hskip 1em plus 0.5em minus 0.4em\relax
  New York, NY, USA: Association for Computing Machinery, Nov. 2017, pp.
  389--401.

\bibitem{kellogg_passive_2016}
B.~Kellogg, V.~Talla, S.~Gollakota, and J.~R. Smith, ``Passive {Wi-Fi}:
  Bringing low power to {Wi-Fi} transmissions,'' pp. 151--164, Mar. 2016.

\bibitem{kellogg_wi-fi_2014}
B.~Kellogg, A.~Parks, S.~Gollakota, J.~R. Smith, and D.~Wetherall, ``{Wi-Fi}
  backscatter: {I}nternet connectivity for {RF}-powered devices,'' \emph{ACM
  SIGCOMM Computer Communication Review}, vol.~44, no.~4, pp. 607--618, Aug.
  2014.

\bibitem{talla_lora_2017}
V.~Talla, M.~Hessar, B.~Kellogg, A.~Najafi, J.~R. Smith, and S.~Gollakota,
  ``{LoRa} backscatter: Enabling the vision of ubiquitous connectivity,''
  \emph{Proceedings of the ACM on Interactive, Mobile, Wearable and Ubiquitous
  Technologies}, vol.~1, no.~3, pp. 105:1--105:24, Sep. 2017.

\bibitem{bharadia_backfi_2015}
D.~Bharadia, K.~R. Joshi, M.~Kotaru, and S.~Katti,
  ``\BIBforeignlanguage{en}{{BackFi}: High throughput {WiFi} backscatter},'' in
  \emph{\BIBforeignlanguage{en}{Proceedings of the 2015 {ACM} {Conference} on
  {Special} {Interest} {Group} on {Data} {Communication}}}.\hskip 1em plus
  0.5em minus 0.4em\relax London United Kingdom: ACM, Aug. 2015, pp. 283--296.

\bibitem{yuanEnablingNativeWiFi2023}
L.~Yuan and W.~Gong, ``Enabling native {WiFi} connectivity for ambient
  backscatter,'' in \emph{Proceedings of the 21st {{Annual International
  Conference}} on {{Mobile Systems}}, {{Applications}} and {{Services}}}, ser.
  {{MobiSys}} '23.\hskip 1em plus 0.5em minus 0.4em\relax Association for
  Computing Machinery, pp. 423--435.

\bibitem{kaplan_direct_2024}
A.~Kaplan, J.~Vieira, and E.~G. Larsson, ``Direct {Link} {Interference}
  {Suppression} for {Bistatic} {Backscatter} {Communication} in {Distributed}
  {MIMO},'' \emph{IEEE Transactions on Wireless Communications}, vol.~23,
  no.~2, pp. 1024--1036, Feb. 2024.

\bibitem{kaplan_dynamic_2022}
------, ``\BIBforeignlanguage{en}{Dynamic {Range} {Improvement} in {Bistatic}
  {Backscatter} {Communication} {Using} {Distributed} {MIMO}},'' in
  \emph{\BIBforeignlanguage{en}{{GLOBECOM} 2022 - 2022 {IEEE} {Global}
  {Communications} {Conference}}}.\hskip 1em plus 0.5em minus 0.4em\relax Rio
  de Janeiro, Brazil: IEEE, Dec. 2022, pp. 2486--2492.

\bibitem{hessar_netscatter_2019}
M.~Hessar, A.~Najafi, and S.~Gollakota, ``Netscatter: enabling large-scale
  backscatter networks,'' in \emph{Proceedings of the 16th USENIX Conference on
  Networked Systems Design and Implementation}, ser. NSDI'19.\hskip 1em plus
  0.5em minus 0.4em\relax USA: USENIX Association, 2019, p. 271–283.

\bibitem{mishraMultiTagBackscatteringMIMO2019}
D.~Mishra and E.~G. Larsson, ``Multi-{{Tag Backscattering}} to {{MIMO Reader}}:
  {{Channel Estimation}} and {{Throughput Fairness}},'' \emph{IEEE Transactions
  on Wireless Communications}, vol.~18, no.~12, pp. 5584--5599, Dec. 2019.

\bibitem{mishraSumThroughputMaximization2019a}
------, ``Sum {{Throughput Maximization}} in {{Multi-Tag Backscattering}} to
  {{Multiantenna Reader}},'' \emph{IEEE Transactions on Communications},
  vol.~67, no.~8, pp. 5689--5705, Aug. 2019.

\bibitem{alim_backscatter_2017}
M.~A. Alim, S.~Saruwatari, and T.~Watanabe, ``Backscatter {MAC} protocol for
  future {Internet} of {Things} networks,'' in \emph{2017 {IEEE} 13th
  {International} {Conference} on {Wireless} and {Mobile} {Computing},
  {Networking} and {Communications} ({WiMob})}, Oct. 2017, pp. 1--7.

\bibitem{choi_devices_2021}
E.-J. Choi, K.~M. Kim, and T.-J. Lee, ``Devices and backscatter tag {MAC}
  protocol for an integrated wireless network,'' in \emph{{International}
  {Conference} on {Information} {Networking} ({ICOIN})}, Jan. 2021, pp.
  209--212.

\bibitem{cao_distributed_2020}
X.~Cao, Z.~Song, B.~Yang, M.~A. Elmossallamy, L.~Qian, and Z.~Han, ``A
  distributed ambient backscatter {MAC} protocol for internet-of-things
  networks,'' \emph{IEEE Internet of Things Journal}, vol.~7, no.~2, pp.
  1488--1501, Feb. 2020.

\bibitem{ma_design_2019}
Z.~Ma, L.~Feng, and F.~Xu, ``Design and analysis of a distributed and
  demand-based backscatter {MAC} protocol for internet of things networks,''
  \emph{IEEE Internet of Things Journal}, vol.~6, no.~1, pp. 1246--1256, Feb.
  2019.

\bibitem{kwon_dominant_2016}
J.-H. Kwon, H.-H. Lee, Y.~Lim, and E.-J. Kim,
  ``\BIBforeignlanguage{en}{Dominant channel occupancy for {Wi}-{Fi}
  backscatter uplink in industrial internet of things},''
  \emph{\BIBforeignlanguage{en}{Applied Sciences}}, vol.~6, no.~12, p. 427,
  Dec. 2016.

\bibitem{wangSpraySpectrumefficientAgile2024}
S.~Wang, Y.~Yan, Y.~Chen, P.~Yang, and X.-Y. Li, ``Spray: {{A
  Spectrum-efficient}} and {{Agile Concurrent Backscatter System}},'' vol.~20,
  no.~2, pp. 42:1--42:21.

\bibitem{valentini_cross-layer_2020}
R.~Valentini, P.~D. Marco, R.~Alesii, and F.~Santucci, ``Cross-layer analysis
  of distributed passive {RFID} systems over faded backscattering {Links},'' in
  \emph{2020 {IEEE} {Wireless} {Communications} and {Networking} {Conference}
  ({WCNC})}, May 2020, pp. 1--6.

\bibitem{koizumi_asynchronous_2023}
R.~Koizumi, Y.~Konishi, K.~Kizaki, T.~Fujihashi, S.~Saruwatari, and
  T.~Watanabe, ``Asynchronous {MAC} protocol for receiver-less backscatter
  tag,'' in \emph{2023 {IEEE} 34th {Annual} {International} {Symposium} on
  {Personal}, {Indoor} and {Mobile} {Radio} {Communications} ({PIMRC})}, Sep.
  2023, pp. 1--6.

\bibitem{sen_successive_2010}
S.~Sen, N.~Santhapuri, R.~R. Choudhury, and S.~Nelakuditi, ``Successive
  interference cancellation: a back-of-the-envelope perspective,'' in
  \emph{Proceedings of the 9th {ACM} {SIGCOMM} {Workshop} on {Hot} {Topics} in
  {Networks}}, ser. Hotnets-{IX}.\hskip 1em plus 0.5em minus 0.4em\relax New
  York, NY, USA: Association for Computing Machinery, Oct. 2010, pp. 1--6.

\bibitem{rezaei_coding_2023}
F.~Rezaei, D.~Galappaththige, C.~Tellambura, and S.~Herath, ``Coding
  {Techniques} for {Backscatter} {Communications}—{A} {Contemporary}
  {Survey},'' \emph{IEEE Communications Surveys \& Tutorials}, vol.~25, no.~2,
  pp. 1020--1058, 2023.

\bibitem{mahdavifar_coding_2015}
H.~Mahdavifar and A.~Vardy, ``Coding for tag collision recovery,'' in
  \emph{2015 {IEEE} {International} {Conference} on {RFID} ({RFID})}, Apr.
  2015, pp. 9--16.

\bibitem{deavours_application_2009}
\BIBentryALTinterwordspacing
D.~D. Deavours, ``\BIBforeignlanguage{en}{Application of passive {UHF} {RFID}
  in intermodal facilities},'' Tech. Rep., Jul. 2009. [Online]. Available:
  \url{https://www.ittc.ku.edu/publications/documents/Deavours2009_41420-14.pdf}
\BIBentrySTDinterwordspacing

\end{thebibliography}

	\begin{IEEEbiography}[{\includegraphics[width=1in,height=1.25in,clip,keepaspectratio]{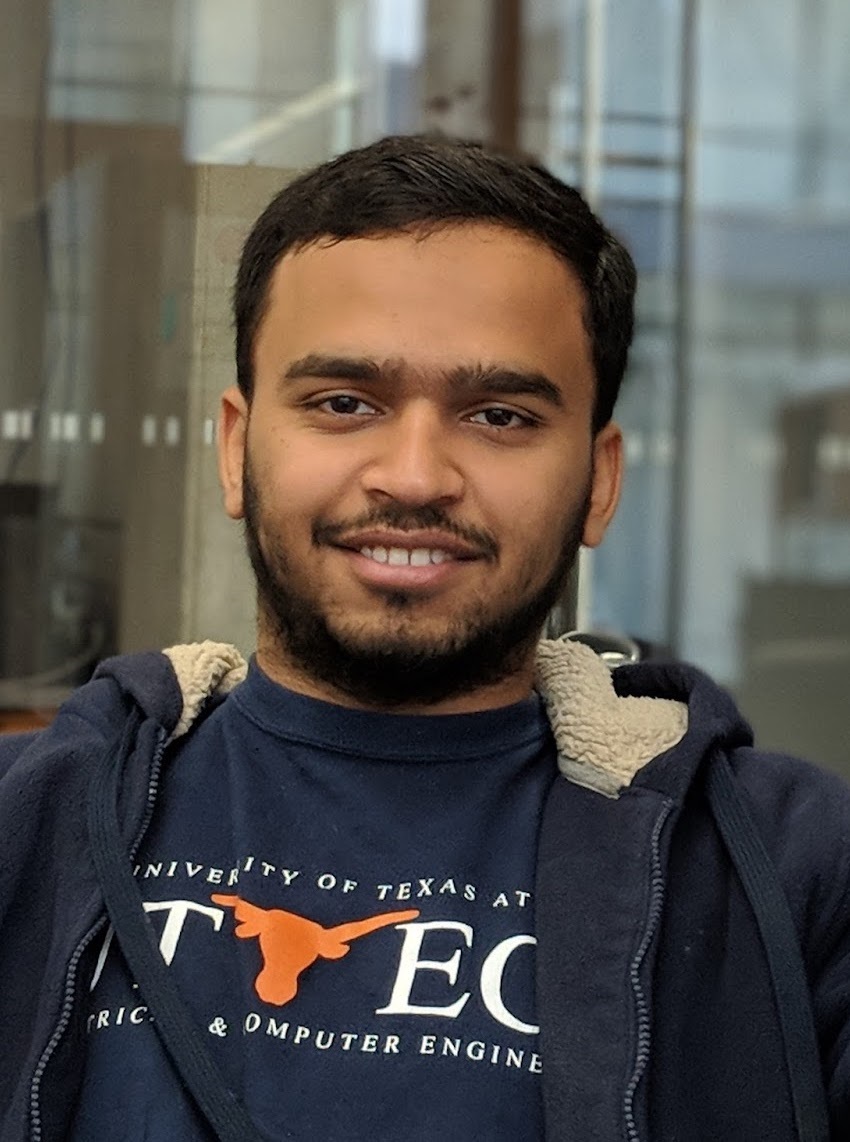}}]{Kartik Patel}
		(S'15 - M'24) received the B.Tech. degree in electronics and communication engineering from the Indian Institute of Technology Roorkee in 2017, and the M.S. and Ph.D. degrees in electrical and computer engineering from the University of Texas at Austin in 2020 and 2024. He is currently Postdoctoral Researcher at Nokia Bell Labs, NJ, USA. He has held internship positions with the Indian Institute of Science Bengaluru; Cisco Innovation Labs, CA, USA; Qualcomm Inc., CA, USA; and Nokia Bell Labs, NJ, USA. His research interests lie at the intersection of wireless networks, sensing, and machine learning with an active focus on system-level validation. Kartik was a finalist for the Best Paper Award at ACM MobiHoc 2019 and co-winner of third place in the Graduate Category of the ACM Student Research Competition at ACM MobiCom 2024.
	\end{IEEEbiography}
	
	\begin{IEEEbiography}[{\includegraphics[width=1in,height=1.25in,clip,keepaspectratio]{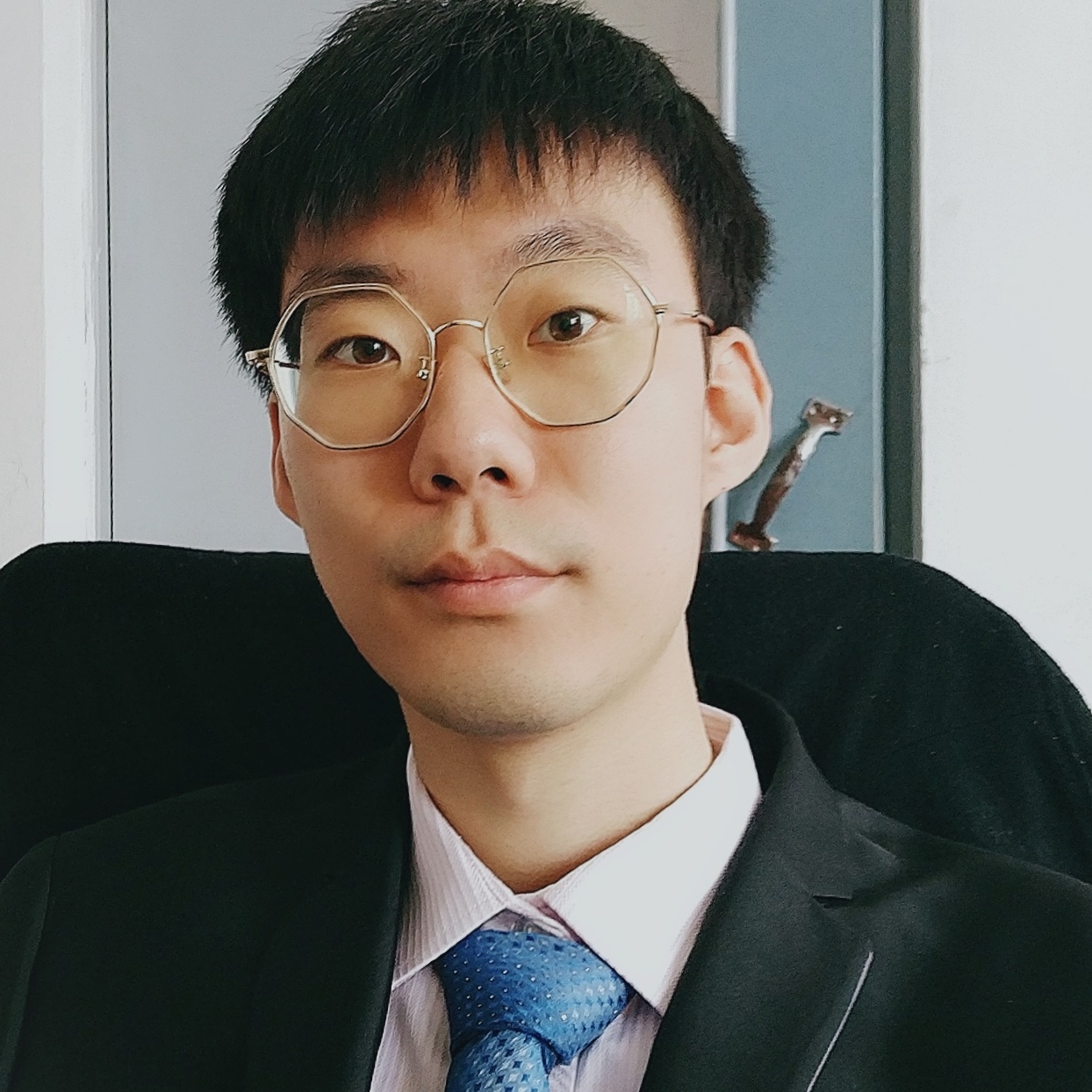}}]{Junbo Zhang}
	    received his Ph.D. degree in electrical and computer engineering from Carnegie Mellon University, Pittsburgh, US in 2024. His research interests lie broadly in building wireless systems for sensing and communication applications, where he has explored a variety of wireless technologies with a specific focus on wireless backscatter networks. His Ph.D. thesis focused on enabling next-generation sensing with RF backscatter and flexible materials.
	\end{IEEEbiography}
	
	\begin{IEEEbiography}[{\includegraphics[width=1in,height=1.25in,clip,keepaspectratio]{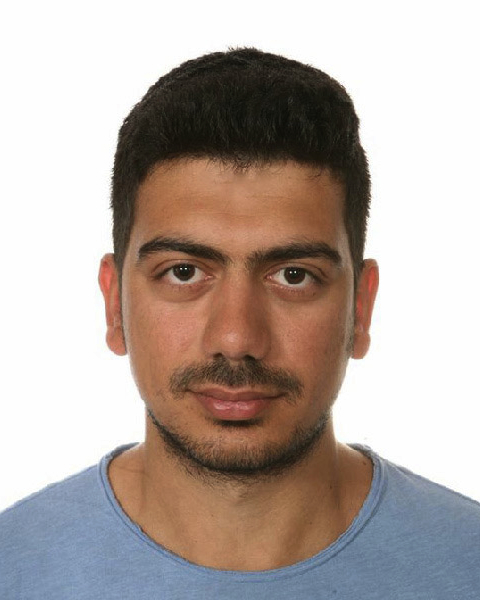}}]{John (Ioannis) Kimionis}
		(S'10–M'17) is a Senior Research Scientist with Nokia Bell Labs in Murray Hill, NJ, USA, where he serves as the technical lead for low-power IoT technologies research, integrating RF front-ends, physical layer processing, and network architectures. He has a track record of over 50 journal and conference papers, magazine articles, and book chapters on spectral- and energy-efficient backscatter radio and RFID, additive manufacturing technologies for low-cost electronics, and mmWave systems.
			
		He received the Diploma of Electronic and Computer Engineering from the Technical University of Crete (TUC), Chania, Greece. He holds a Master of Science degree from the same school, and a Master of Science from the School of Electrical and Computer Engineering, Georgia Institute of Technology (GaTech), Atlanta, GA, USA. He obtained the Ph.D. degree from Georgia Institute of Technology in 2017, where he had been a Research Assistant with the ATHENA Group. In 2018, he joined Nokia Bell Labs in Murray Hill, NJ, USA as a full-time research scientist.
			
		Dr. Kimionis was a recipient of fellowship awards for his undergraduate and graduate studies at TUC, participated in NSF and DTRA-funded projects during his PhD work, and was a Texas Instruments Scholar for his mentoring service for the Opportunity Research Scholars Program of GaTech. He has been a co-recipient of Best Student Paper Awards at the IEEE International Conference on RFID-TA in 2011 and 2014, the third Bell Labs Prize Award for game-changing technologies on printed electronics and low-cost communications in 2016, the Best Industry Paper Award at IEEE RFIC Symposium in 2020, and has been invited to contribute several articles and book chapters in his research fields. He served as a Finance Chair of the IEEE International Conference on RFID in 2018 and serves as a Board member of the IEEE MTT TC-26 RFID, Wireless Sensor, and IoT Committee.
	\end{IEEEbiography}

	\begin{IEEEbiography}[{\includegraphics[width=1in,height=1.25in,keepaspectratio]{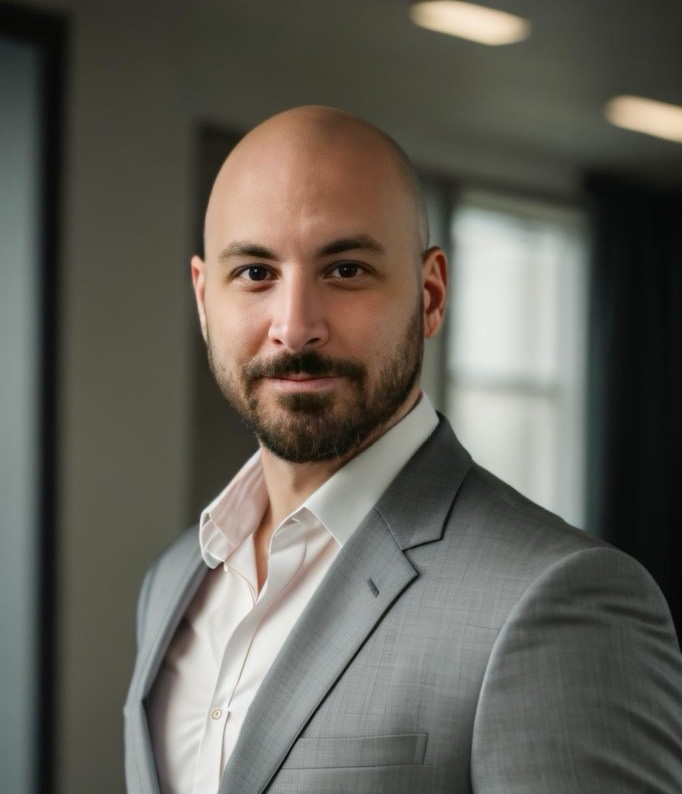}}]{Lefteris (Lef) Kampianakis}
	    (S'15) received the B.S. and M.S. in Electronic and Computer Engineering from the Technical University of Crete, Greece. His diploma thesis on over-the-air programmable wireless sensor networks (WSN) received the best diploma thesis award on the pan-hellenic IEEE thesis competition for the years 2009-2011. During his M.S. he developed low-cost and low-power WSNs for environmental sensing exploiting backscatter communication principles. He received his Ph.D from the Department of Electrical Engineering at the University of Washington where he developed high data-rate ultra low-power backscatter communication systems for brain-computer interfaces. His PhD work was nominated for the James Yang outstanding doctoral student award (top 3). His research work has been published in over 15 peer-reviewed journals and conferences. From 2018-2022, he worked as a Senior Electrical Engineer at the RnD team of Cirtec Medical where he was developing novel Class III implantable medical systems for neural sensing and neuro-modulation. In June 2022, he joined the Nokia Bell Labs Data and Devices team as a Backscatter Devices Researcher where he is one of the tech-leads for development of ultra low power backscatter communication systems for sensing and localization. His research interests include backscatter communication, medical devices and wireless sensor networks. His hobbies include powerlifting, and Greek dancing.
	\end{IEEEbiography}
	
	\begin{IEEEbiography}[{\includegraphics[width=1in,height=1.25in,keepaspectratio]{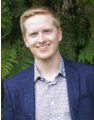}}]{Michael S. Eggleston}
		received the B.S. degree in electrical engineering and physics from Iowa State University, IA, USA, and the Ph.D. degree in electrical engineering from the University of California, Berkeley, CA, USA. In 2015, he joined Nokia Bell Labs, Murray Hill, NJ, USA, where he currently leads the Data and Devices Research Department. This interdisciplinary team is building the devices and AI systems that will seamlessly connect the physical and digital worlds of the future. He holds more than ten patents on integrated photonic and sensing technologies. His research interests include non-invasive human sensing, energy-autonomous devices, and quantum technologies.
	\end{IEEEbiography}
	
	\begin{IEEEbiography}[{\includegraphics[width=1in,height=1.25in,clip,keepaspectratio]{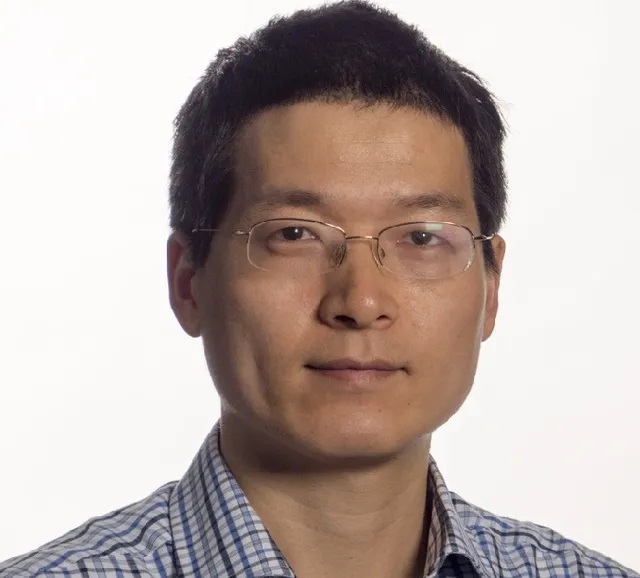}}]{Jinfeng Du}
 	    (Senior Member, IEEE) is leading the Department of Radio Systems USA in Nokia Bell Labs. He received the B.Eng. degree in electronic information engineering from the University of Science and Technology of China (USTC), Hefei, China, in 2004, and the M.Sc., Tekn. Lic., and Ph.D. degrees from the Royal Institute of Technology (KTH), Stockholm, Sweden, in 2006, 2008, and 2012, respectively. He was a Postdoctoral Researcher with the Massachusetts Institute of Technology (MIT), Cambridge, MA, USA, from 2013 to 2015. Since joining Bell Labs in 2015, he has focused on fundamentals and disruptive concepts for wireless communications, especially in communication theory, radio system design and optimization, millimeter-wave propagation measurements and modeling.
	\end{IEEEbiography}

\end{document}